\documentclass[11pt,a4paper]{article}
\usepackage[hyperref]{acl2020}
\usepackage{times}
\usepackage{graphicx}
\usepackage{latexsym}
\usepackage{etoolbox,lipsum}
\usepackage{mathrsfs}
\usepackage{amsmath}

\usepackage{mwe}

\usepackage{microtype}

\aclfinalcopy 

\usepackage{ amssymb }
\usepackage{ mathrsfs }

\title{ Enhancing Audio Perception of Music By AI Picked Room Acoustics\\
}

\author{Prateek Verma and Jonathan Berger\\
  Stanford University\\
  Stanford, CA, USA\\
  \texttt{prateekv@stanford.edu, brg@ccrma.stanford.edu } }

\usepackage{cuted}
\usepackage{capt-of}

\begin{document}
 \maketitle
 \begin{strip}\centering
\includegraphics[width=\textwidth]{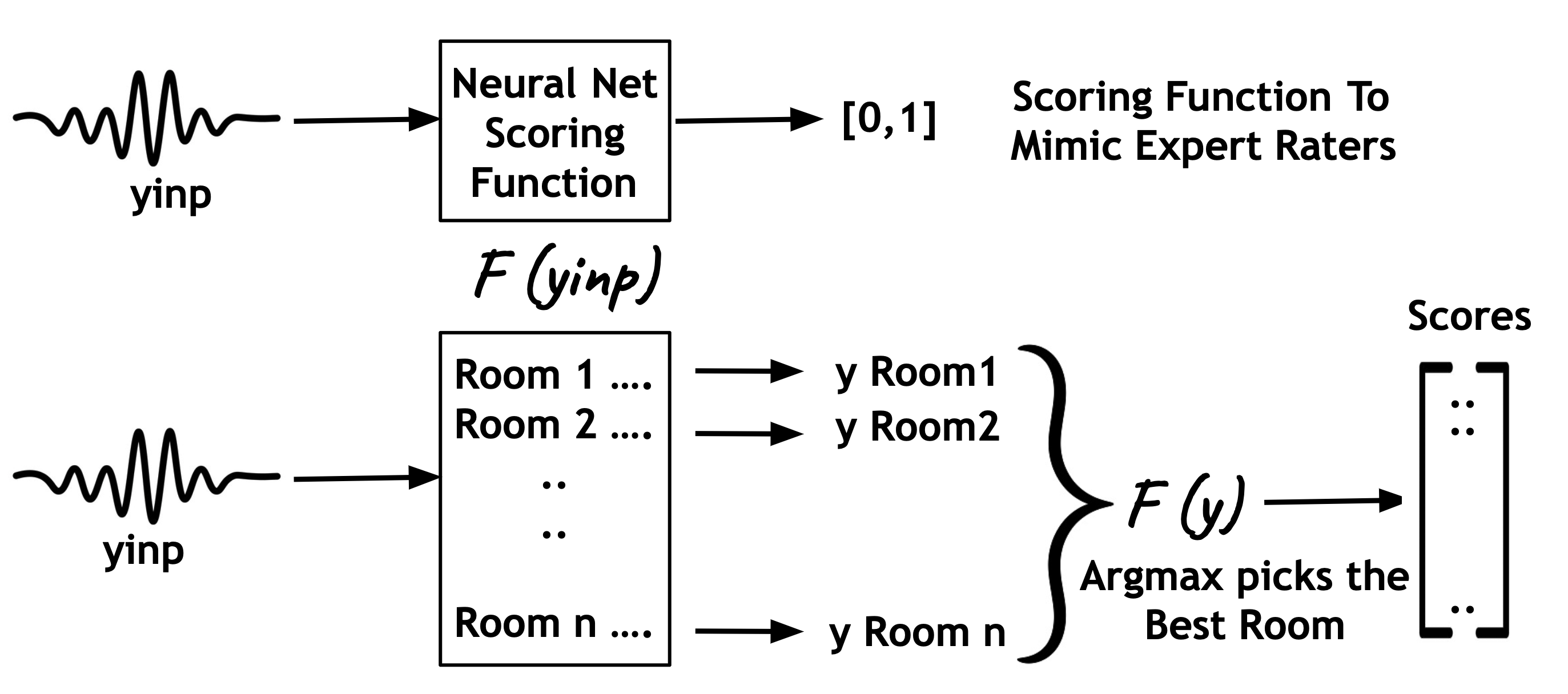}
\captionof{figure}{Proposed method in the paper. We use room acoustics as the only transform to enhance quality of a musical note, due to their ability to change the volume and timbre dynamics of the signal.
\label{fig:feature-graphic}}
\end{strip}
 
\begin{abstract}
Every sound that we hear is the result of successive convolutional operations (e.g. room acoustics, microphone characteristics, resonant properties of the instrument itself, not to mention characteristics and limitations of the sound reproduction system). In this work we seek to determine the best room in which to perform a particular piece using AI. Additionally, we use room acoustics as a way to enhance the perceptual qualities of a given sound. Historically, rooms (particularly Churches and concert halls) were designed to host and serve specific musical functions. In some cases the architectural acoustical qualities enhanced the music performed there. We try to mimic this, as a first step, by designating room impulse responses that would correlate to producing enhanced sound quality for particular music. A convolutional architecture is first trained to take in an audio sample and mimic the ratings of experts with about 78 \% accuracy for various instrument families and notes for perceptual qualities. This gives us a scoring function for any audio sample which can rate the perceptual pleasantness of a note automatically. Now, via a library of about 60,000 synthetic impulse responses mimicking all kinds of room, materials, etc, we use a simple convolution operation, to transform the sound as if it was played in a particular room. The perceptual evaluator is used to rank the musical sounds, and yield the "best room or the concert hall" to play a sound. As a byproduct it can also use room acoustics to turn a poor quality sound into a "good" sound. 
\end{abstract}

\section{Introduction and Related Work}

Humans routinely encounter a wide range of sounds. These ubiquitous signals range from speech and other humanly produced sounds, to sounds of other organisms, to sounds created by atmospheric interaction with the environment, to mechanically and artificially produced sounds \cite{gemmeke2017audio}. Sound enables communication, engages, and promotes understanding of the world around us. The sounds that we hear from this rich platter are typically transformed by reflections and resonances as the signal interacts with objects and structures. The resonant features of musical instruments, and the directionality of signal propagation affect the sound's timbral characteristics and intensity.
\newline
Humans also tend to favor sounds whose timbre is enriched by reverberation over sounds produced and recorded under anechoic conditions. Although the reasons for such preference are not entirely clear, studies that explore the acoustics of ancient caves inhabited by paleolithic humans suggest that the reverberant qualities of these caves might have functioned as the first auditoriums in which humans experienced sound\cite{fazenda2017cave}. It is conceivable that as early humans found safety and comfort inhabiting caves, they also found comfort in their dwelling's rich acoustical environment\cite{shipton201878}. Furthermore, this may suggest an evolutionary source of human preference for rich reverberant environments. As music developed, vocal sounds, performance practices, and musical instruments adjusted to the acoustical characteristics of the structures in which they were sounded and heard. Listening studies show a preference for reverberated sound. Humans also adjust their playing according to the environment (acoustic space) they are in. Expert performers adjust, often with little or no conscious awareness, how they execute a sound based on the architectural acoustical features. Musicians dislike playing in anechoic environments noting that "the lack of reflection or reverberation is acoustically stifling affecting their playing styles" \cite{maconie2010chapter}. Thus in this work, we allow a neural architecture to determine the optimal room (through the room's impulse response) in which to play a musical sound.
\newline
Neural architectures have revolutionized how people approach traditional signal processing ideas, with advancements in music synthesis via generative modelling\cite{dhariwal2020jukebox,verma2021generative}, audio understanding \cite{verma2021audio} to name a few. With these advancements, neural architectures have become powerful in understanding contents, and synthesis of the audio signals in problems such as speech recognition \cite{baevski2020wav2vec} and audio synthesis \cite{wang2017tacotron}. In the recent literature, neural architectures have been used to evaluate the quality of sound of interest. \cite{verma2019learning} first used a neural architecture to evaluate the sounds of a poor-quality cello, often used by amateur musicians with that of a high-quality cello, often used by expert players. By successfully allowing a neural architecture to rate the quality of the sound, they then used it as a loss function to map a cello sound from poor quality to high quality without having access to an aligned dataset. In some sense, this work is similar with the only difference being that here it is adapted to room impulse responses. There have been other works like using neural architectures to predict mean opinion scores directly e.g. to improve speech synthesis \cite{lo2019mosnet}. \cite{slaney2020auditory} came up with measures that can understand the perceptual qualities of speech signals. The ideas in these works are similar: Humans use some rules to rate audio samples, and a neural architecture can easily learn and mimic these features. The contributions of this work are as follows:

i) To the best of our knowledge across literature we are the first ones to use impulse response as a way to enhance the perceptual qualities of music signal (audio signal)

ii) We learn a perceptual score for the task of our interest using neural architecture and use it to evaluate musical signals that would be transformed using room impulse response. This in a way used correctly, can be used to enhance the perceptual score of a musical signal, by using the room acoustics itself as a perception enhancer filter.

The organization of the paper is as follows: Section 1 introduces the problem with related work. Section 2 describes the dataset used in our work along with the methodology used in section 3. We follow it up with results and Section 5 describing the conclusion and future work.

\section{Dataset} For simulation of various playing conditions, we use impulse responses that we simulated for improving speech recognition in reverberated environments \cite{ko2017study}. Similar to musical acoustics, for speech signals too, room acoustics changes various aspects of audio signals, such as timbre, volume envelops, and attack rate to name a few. The actual room impulse measurements are difficult to acquire, with techniques involving setups but not limited to balloon pops, sweeping sinusoidal signals, etc. \cite{abel2010estimating}. Due to the non-trivial nature of measure room impulse response, we started with using  \cite{ko2017study} as a primary source of IR measurements for our study. To the best of our knowledge, this is the largest possible resource of a collection of impulse responses in the literature that we came across. We also initially experimented with \cite{traer2016statistics}, which were about 271 real-world impulse responses but did not use them for our study due to the sheer diversity, richness, and size of the IRs generated by \cite{ko2017study}. In speech signals, impulse response measurements are primarily used as a data augmentation tool to enhance the performance, and robustness of speech recognition algorithms \cite{yoshioka2012making}. There are in total about 200 rooms uniformly distributed between small (1-10m), medium (10-30m), and large (30-50m) rooms. For each of the 600 rooms, the height is uniformly sampled from 2-5m with different absorption coefficients and different locations inside a room for every IR. This yields a diverse set of room acoustics.

To evaluate the perception of the sounds, we use recorded individual musical notes as described in \cite{romani2015real}. To acquire a consensus of what constitutes a 'good' versus a 'bad' musical sound, we ask professional performers to rate a collection of single notes in terms of performance quality describing attributes of pitch stability, timbre stability, richness, and attack. As each of these attributes are influenced by characteristics of the room acoustic we show that room acoustics alter the overall perception of musical notes. We choose three instruments, clarinet, flute, and trumpet, for a total of about 2000 notes. A neural architecture is used to mimic the function of human raters via convolutional architecture making it "human-inspired neural scoring function" to evaluate any sound of interest, in our case that passes through a room impulse convolution operation to simulate as if the note was played in a particular room.

\section{Methodology and Setup} 
This section describes the methodology used by us, in the proposed work, in three sections. Firstly how to mimic room acoustic via synthetic impulse response, followed by a neural architecture that was used to rate the output of impulse responses.

\subsection{Mimicking Room Acoustics via Synthetic Impulse Response} 
For the case of our study, we assume that there is no presence of noise (arising from outside sources or concerning imperfections in the acoustic measurement/noisy impulse responses/measurement noise from room impulse measurement). This can encompass a wide array of settings and can yield simple mathematical modeling and an elegant setup in our case. For the users interested in a systematic simulation of far-field musical sound, they can refer to several techniques one of them being \cite{chechile2017vampireverb}. Reverberation using signal processing algorithms is a field in itself, and \cite{valimaki2016more} describes three main categories of how to carry out reverberation using delay networks, convolution-based methods, and physical models of rooms. For the rest of our work, we would only use a convolution-based method without taking into account the measurement noise and other noises as described in \cite{ko2017study}. Multiple balloon pops were used to create the room impulse responses to recreate the acoustics of Hagia Sophia in Stanford's Bing Concert hall as described by work done \cite{abel2013recreation}. For the current work, we would assume that an impulse response simulated at a location point $p$, in a room $r$ is $h_{pr}[n]$. Then a signal played in an anechoic or reverberated environment $x[n]$ would sound in the room $r$ at point $p$, as $y[n] = x[n] \ast h_{pr}[n]$. For any sound of interest $ x[n]$, we have in total of 60,000 impulse responses.

\subsection{Neural Architecture} We particularly use a convolutional architecture to mimic the perceptual ratings of expert human evaluators. For this, we train small convolutional architectures on 5s of audio. The audio signal sampled at 16kHz was converted into its spectral representation in our case mel-spectrogram, with 10ms hop, and 96 bins yielding a spectrogram representation of shape 96 x 500 corresponding to the time steps for a 5s audio. This representation is fed to a 5-layer convolutional architecture with the following setup: We use 7x7 filters with strides of 5, with 16 filters in the first three layers. The dropout rate \cite{JMLR:v15:srivastava14a} was fixed to be 0.5 to heavily regularize these small architectures.  The number of filters is small, due to about 2000 notes present in the training corpus that has been rated by more than 3 musicians. The output of the last layer is flattened followed by a dense layer of size 256 followed by a single neuron that gives a score between 0 and 1, similar to \cite{verma2019learning}. We choose a score of 0 to represent poor sound quality, and a score of 1 to represent a sound perceived as 'musical'. We only use labels that are agreed upon by 2 or more raters. This is validated on about 200 sounds. The loss function chosen to be Huber Loss \cite{huber1992robust} between the predicted score and the actual rating, primarily to make training robust to outliers, that can perhaps arise from human raters. 

We use heavy data-augmentation for musical signals like random volume scaling, random cut-outs, stripping timbral content from spectral, backbone \cite{verma2020framework},  flips (both vertical and horizontal), and adding random noise to improve the robustness of our predictions \cite{mcfee2015software}. These techniques are used in music signal understanding for applications such as audio understanding \cite{verma2021audio,hershey2017cnn} This is primarily done in our setup for two reasons. Firstly, they improve the performance of the task of interest.
Secondly, and more importantly, we have a small high-quality sample of professional musicians as human raters to depend on a variety of factors and not enable the neural architecture to latch onto one-two salient feature. 

All architectures were trained for about 50 epochs with validation loss used to come up with the best architecture. Adam optimizer \cite{kingma2014adam} was used with a learning rate of 1e-4 and reduced till 1e-6 or earlier whenever validation loss does not start decreasing. We used Tensorflow \cite{abadi2016tensorflow} framework for all of our experiments.

\subsection{Connecting Perceptual Scores with Room Acoustics for Ranking } With the library of impulse responses simulated for a wide array of room sizes, materials, geometries, and point sources in a single room, we use all 60,000 impulse responses to simulate how a musical note would sound in a particular acoustic environment. To the best of our knowledge, no such study if musical context involving different room exists. 
We take musical notes, and create 60k copies of them, mimicking the output of various room acoustics via a simple convolution operation of the musical signal with that of the impulse response of the simulated room. We then use our scoring function to evaluate each of the transformed audio signals, on the attributes learned by neural architecture and use it to pick the best room in which a note should be played. This can also be used to enhance and improve the quality of a sound through the interaction of the acoustical qualities of the space and the sound, and increase the saliency of the musical feature of interest.

\section{Results and Discussion} 
This section shows how the room acoustics alter the timbral and the volume dynamics to alter the characteristics of a given note through some quantitative examples. 
\subsection{Altering the energy dynamics}
Energy dynamics play a major role in defining perceptual qualities of a given note. \cite{romani2015real} had asked experienced musical raters to describe the characteristics on which they judged, and the timbral characteristics along with energy dynamics played an important role. We through an example here see how the room acoustics was able to alter the energy dynamics. Figure 2 describes a single musical note with little to no presence of energy modulation, sometimes thought of as a feature of aesthetics and exhibiting the overall skill of a musician. This 5s note was initially given a poor score ( less than 0.1) and boosted to be a high score (greater than 0.9), by altering the dynamics of the energy variations.

\begin{figure}[ht]
    \centering
    \includegraphics[width=\linewidth]{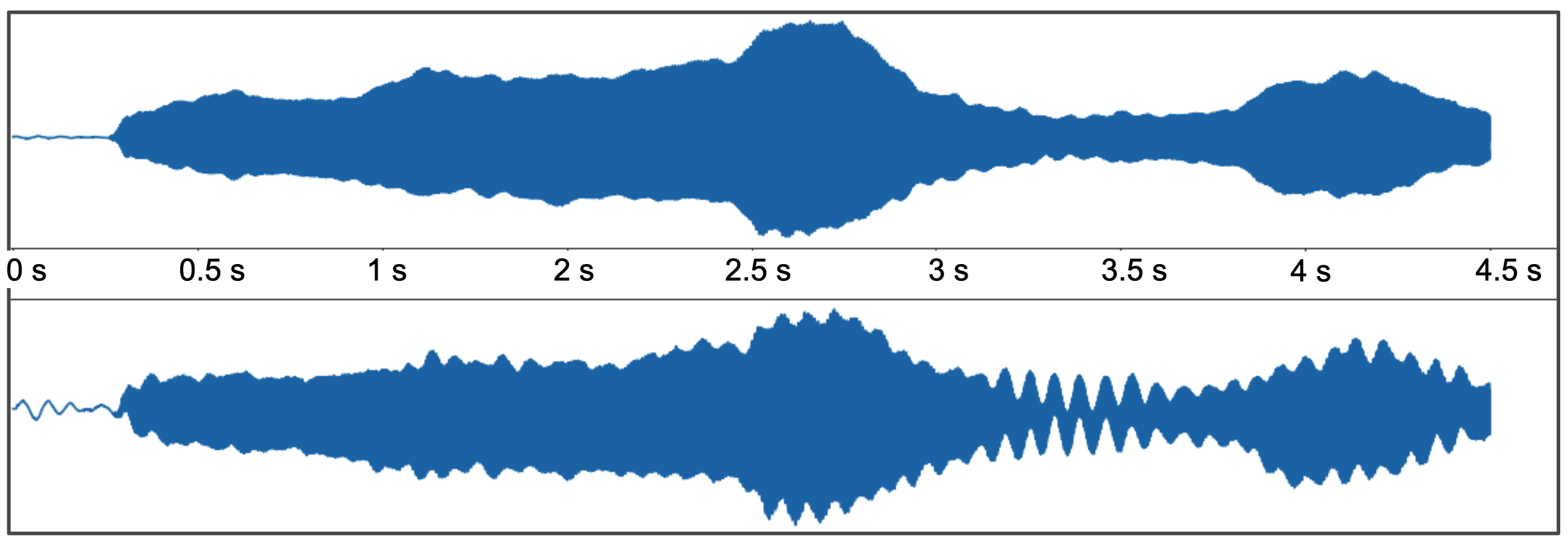}
    \caption{Figure describing a poor note played by a musician (above) and the same note below that was boosted by a room impulse response, as if it was played in a particular acoustic space. It was chosen by our algorithm to maximize the score to improve the quality of the note(below). Notice how energy modulations are present which is considered an embellishment, often ignored in a low-quality sound.   
    }
    \vspace{-15pt}
\end{figure}

\subsection{Altering timbre} We via the same example show firstly how the timbre of the note changes. This is just for the sake of demonstration and such behaviour was observed in a wide-array of sounds. We can see in Fig. 3 as to how on left the spectral centroid changes, results in a higher, richer tone as compared to the original sound. Richer sounds are associated with bright, more complex enhanced tones, as compared to lower frequency tones. On the right, we show the optimal room impulse response. One could see how there is a repetitive structure present in the impulse response that can add structure, and energy modulations to the energy envelop that is absent in the note without the interaction with the room acoustics.

\begin{figure}[ht]
    \centering
    \includegraphics[width=\linewidth]{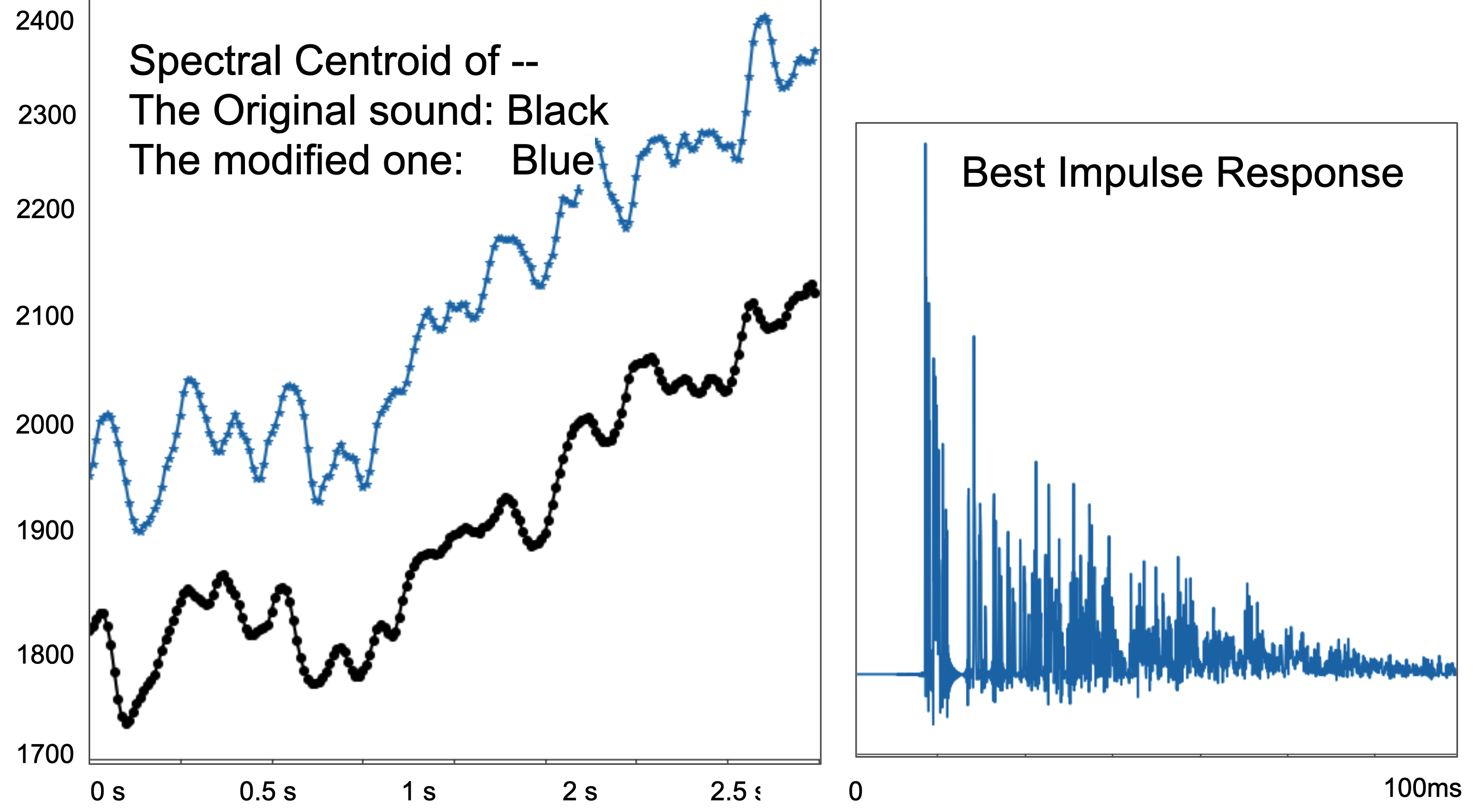}
    \caption{Figure on the left showing the spectral centroid and on the right the room impulse response that enhanced the brightness of the sound
    }
    \vspace{-15pt}
\end{figure}

\subsection{Demonstrating Improvement Statistics of Notes}

We through an example, showed how we could alter a single musical note, both in timbre and energy dynamics. In order to show these behaviour across the dataset, we design the following experiment. For 250 notes from a held out validation set only for the low-quality note, we compute the distribution of the scores, and take only those that are below 0.5 or are classified by our algorithm to be of a poor quality. Next, we run it through our dataset of impulse responses, and find the optimum room acoustics to maximise the quality of the sound and note the value passed through our quality assessment architecture. 
\begin{figure}[ht]
    \centering
    \includegraphics[width=\linewidth]{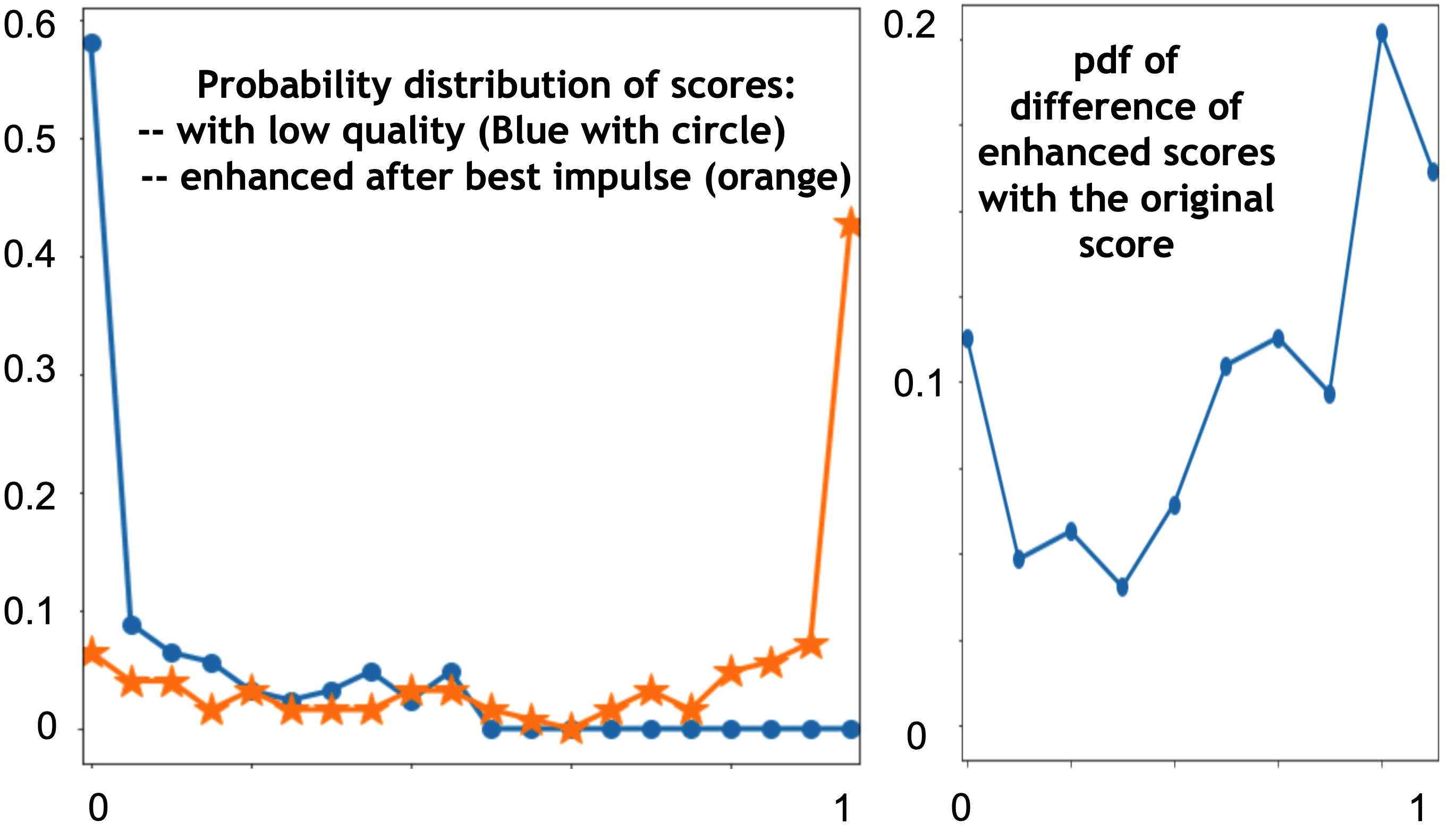}
    \caption{Qualitative results of a sample subset of 250 musical notes, with low scores and their enhanced score computed as a probability distribution (left) and the distribution of the difference of the scores(right). The x-axis denotes the actual scores on left and the difference of the scores on the right. 
    }
    \vspace{-15pt}
\end{figure}

The value (of the enhanced note) now is recorded. For demonstration purposes, we compute the histogram statistics of the following as seen in Figure 4.i) The distribution of the quality of the notes less than 0.5 and the distribution of the notes that are enhanced by picking the best room, and scoring it. ii) The distribution of the difference of the scores between the enhanced output and the initial output. Both of these scores in i) and ii) would lie between [0,1]. We again emphasize that we are using purely room acoustics as a filter to achieve it. We see from Figure 4, that we are able to enhance the score of a majority of sounds, although there still exist a small percentage of notes, that could not be improved upon. This is further validated by looking at the distribution of the difference of the enhanced scores with that of the poor quality notes.

\section{Conclusion and Future Work} 
We have successfully demonstrated a pipeline capable of enhancing the perception of a sound with modeled room acoustics by using a large-scale library of recorded/simulated room impulse responses. Using a neural architecture to understand the perceptual properties of a sound, by mimicking human preference ratings we build a neural evaluator that can score an audio signal. This scoring mechanism is used to evaluate a musical sound played in thousands of different acoustic environments by using room impulse responses in our library. The best room acoustics is chosen according to the characteristics of the audio signal, which results in the highest score from the neural evaluator. Doing large scale listening experiments, is something that we have in our pipeline to further validate the current results. Extending this concept beyond musical notes is vastly interesting, and can be useful in a wide spectrum of applications, particularly in speech, with widespread adoption of smart speakers.

\section{Acknowledgements}
This work was supported by the Institute of Human-Centered AI at Stanford University (Stanford HAI) through a generous Google Cloud computing grant for the academic year 2021-22 to carry out computational experiments. We thank both the HAI as well as Google for the initiative. This work was supported in part by a grant from the Templeton Religion. We would like to thank Prof. Chris Chafe for his useful comments about this work. Prateek Verma would also like to thank Robert Crown Library at SLS, and its staff for providing such good workspaces while this work was done during the height of the Omicron surge in Winter 2022.
\bibliography{acl2020}
\bibliographystyle{acl_natbib}
\end{document}